\documentclass[aps,twocolumn,showpacs,prb]{revtex4}
\usepackage{graphicx}
\usepackage{dcolumn}
\usepackage{amsmath}
\begin{document}
\title{Multiphoton processes in microwave photoresistance of 2D electron system}
\author{M. A. Zudov$^{1,2}$}
\author{R. R. Du$^{1}$}
\affiliation{
$^1$Department of Physics, University of Utah, Salt Lake City, Utah 84112\\
$^2$School of Physics and Astronomy, University of Minnesota, Minneapolis, Minnesota 55455
}
\author{L. N. Pfeiffer}
\author{K. W. West}
\affiliation{Bell Laboratories, Lucent Technologies, Murray Hill, New Jersey 07974}
\begin{abstract}
We extend our studies of microwave photoresistance of ultra-high mobility two-dimensional electron system (2DES) into the high-intensity, non-linear regime employing both monochromatic and bichromatic radiation.
Under high-intensity monochromatic radiation $\omega$ we observe new zero-resistance states (ZRS) which correspond to rational values of $\varepsilon=\omega/\omega_C$ ($\omega_C$ is the cyclotron frequency) and can be associated with multiphoton processes.
Under bichromatic radiation $\omega_1,\omega_2$ we discover new resistance minimum, possibly a precursor of bichromatic ZRS, which seems to originate from a frequency mixing process, $\omega_1+\omega_2$. 
These findings indicate that multiphoton processes play important roles in the physics of non-equilibrium transport of microwave-driven 2DES, and suggest new directions for theoretical and experimental studies.
\end{abstract}
\maketitle

Intense current interest to millimeterwave photoconductivity of two-dimensional electron systems (2DES) was triggered by the discovery of microwave-induced resistance oscillations (MIRO)\cite{zudov:201311,ye:2193} and zero-resistance states (ZRS),\cite{mani:646, zudov:046807} emerging in ultra-high mobility samples at weak magnetic fields and low temperatures.
While a great deal of attention has been paid both experimentally,$^{5-19}$
and theoretically$^{7,20-49}$
In particular, some recent experimental results appear to challenge existing theories.\cite{yang:up, smet:116804}
To advance our understanding of the phenomena we explore new experimental regime of high-intensity microwave (MW) radiation.

Phenomenologically, MIRO appear due to oscillatory microwave photoresistance (PR), $\Delta R^\omega = R^\omega-R^0$, which (i) is a periodic function in inverse magnetic field, $1/B$ with a period given by $\varepsilon=\omega/\omega_C$, where $\omega=2\pi f$ and $\omega_C=eB/m^*$ are microwave and cyclotron frequency, respectively, (ii) can take both positive (maxima) and negative (minima) values depending on $\varepsilon$, and (iii) can approach the background dark resistance by absolute value under certain experimental conditions.
Although one could expect appearance of negative resistance at some of the MIRO minima, experiments (and theory\cite{andreev:056803}) usually (see, however Ref.\,10) show that 2DES instead exhibits ZRS, as the resistance saturates close to zero.
Since the resistance maxima ($+$) and minima ($-$) appear in pairs\cite{zudov:041304} situated roughly symmetrically about cyclotron resonance harmonics, i.e., $\varepsilon_j^{\pm} \approx j \mp \phi_j$, they are commonly associated with an {\em integral} $j$ ($\varepsilon_j^{\pm} \approx j$).
All ZRS reported to date were observed at 
\begin{equation}
\varepsilon_j=j+\phi_j,\,\,\,j=1,2,3,...\,
\label{sp}
\end{equation}
where $\phi_j$ is experimentally found to be $0<\phi\le 1/4$.

The work reported here was provoked by peculiar photoresistance features far from integral $\varepsilon$, which were first noticed in experiments on moderate mobility samples.\cite{zudov:201311}
In particular, prominent resistance maximum and minimum were observed about $\varepsilon \approx 1/2$ at $f=45$ GHz.\cite{zudov:201311}
Later experiments confirmed this observation\cite{dorozhkin:577, dorozhkin:201306, willett:026804} and even reported features close to other rational values of $\varepsilon \approx j/m$, e.g. 3/2, 5/2, and 2/3.\cite{zudov:046807,zudov:041304}
It was proposed,\cite{zudov:041304} that these features originate from two-($m=2$) and three-photon ($m=3$) processes; here, an electron absorbs $m$ photons to jump $j$ Landau level spacings.
Moreover, it was noticed that the most readily detectable multiphoton structure at $\varepsilon \approx 1/2$ can develop a peak comparable in magnitude to the integral at $\varepsilon \approx 1$.\cite{zudov:041304}
Therefore, it is natural to ask if appropriate experimental conditions can be reached for multiphoton minima to evolve into ZRS.
New ZRS would appear in accordance with an analog of Eq.\,(\ref{sp}), generalized to incorporate multiphoton processes:\cite{zudov:041304}
\begin{equation}
\varepsilon_j^{(m)}=\frac j m +\phi_j^{(m)}.
\label{mp}
\end{equation}
Here $m$ is the number of photons involved in a process and, therefore, such ZRS can be associated with a {\em ratio} $j/m$ ($\varepsilon_j^{(m)} \approx j/m$).

On general basis, one expects that multiphoton ZRS would require much higher radiation intensities and, as such, would be difficult to detect.
Along this line, significant experimental advantages are offered by lowering microwave frequency.
First of all, lower frequency ensures single-mode operation of the waveguide which guarantees higher local intensity at the sample center. 
Second, lower frequency results in higher photon flux which scales with the inverse of the photon energy. 
Finally, low frequency moves the strongest two-photon feature ($\varepsilon_1^{(2)} =1/2$) out of the strong Shubnikov-de Haas regime where photoresponse becomes relatively weak.\cite{studenikin:245313} 


In this paper we report on transport studies of ultra-clean 2DES under intense microwave irradiation, where the system reveals characteristically new features in photoresistance spectrum.
The most prominent are new zero-resistance states emerging about rational values of $\varepsilon$.
While more pronounced rational ZRS are situated close to half-integers, e.g. two-photon $\varepsilon_1^{(2)} \approx 1/2$, $\varepsilon_3^{(2)} \approx 3/2$, we also observe formation of a three-photon ZRS in the vicinity of $\varepsilon_2^{(3)} \approx 2/3$.
Other interesting observations include reverse power dependence of the resistance peaks, accompanied by peak narrowing and phase reduction, as well as dramatic suppression of the magnetoresistance at $\varepsilon<1/2$.
Finally, under bichromatic (frequencies $\omega_1,\omega_2$) MW radiation we detect a new resistance minimum which seems to originate from a frequency mixing process, $\omega_1+\omega_2$, and possibly presents the first experimental support for bichromatic ZRS.

Our sample was cleaved from a MBE-grown, symmetrically doped Al$_{0.24}$Ga$_{0.76}$As/ GaAs/Al$_{0.24}$Ga$_{0.76}$As quantum well structure.
After brief illumination with visible light at low temperature, electron mobility, $\mu$ and density, $n_e$ saturated at $\approx$\,$2 \times 10^7$ cm$^2$/Vs and $3.6 \times 10^{11}$ cm$^{-2}$, respectively.
Experiment was performed in a top-loading $^3$He refrigerator equipped with a superconducting solenoid.
For monochromatic experiments we employ a lower frequency ($f=27$ GHz) microwave source to ensure single-mode operation of the WR-28 waveguide used to deliver radiation down to the sample.
With microwaves, the temperature of the coolant was in the range between 1.1 K and 1.8 K depending on the microwave intensity.
Using this temperature rise we estimate that the total microwave power at the bottom flange of the waveguide was from 2 to 12 times larger than in our previous ($\sim 100 \mu$W) studies.\cite{zudov:046807} 
At these intensities one can anticipate potential complications associated with rectification of microwave radiation on sample contacts. 
To check against such effects we have verified that measured voltages remain linear with the excitation current and that zero-field voltage does not depend on microwave intensity.
To implement bichromatic MW experiments, EM waves from two Gunn diodes tuned to distinct frequencies ($f_1=\omega_1/2\pi=31$ GHz, $f_2=\omega_2/2\pi=47$ GHz) were combined using a hybrid ``T'' mounted on the top flange of the waveguide.
All the data presented here were recorded using conventional quasi-dc lock-in technique (1 $\mu$A, 17 Hz), at a constant temperature, under continuous MW illumination of fixed frequency and power, while sweeping the magnetic field.



\begin{figure}[t]
\includegraphics{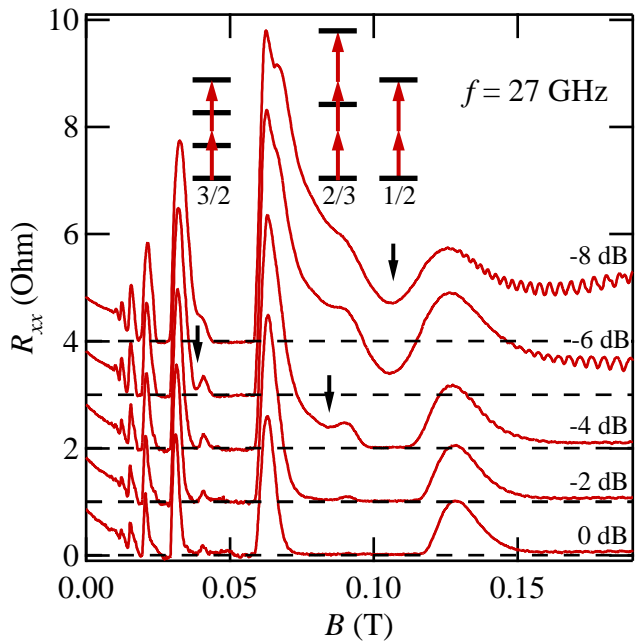}
\caption{[color online]
Magnetoresistance under microwave illumination of $f=\omega/2\pi=27$ GHz, for selected radiation intensities.
Traces are vertically offset for clarity and are marked by attenuation.
Vertical downward arrows mark the positions of oscillations minima corresponding to rational values of $\varepsilon \approx 3/2$, 2/3, and 1/2, from left to right, which develop into new ZRS with increasing intensity.
Insets illustrate virtual multiphoton processes responsible for these features at rational $\varepsilon$.
At $B>0.15$ T non-resonant suppression of resistance is observed.
}
\label{f1}
\end{figure}
\begin{figure}[t]
\includegraphics{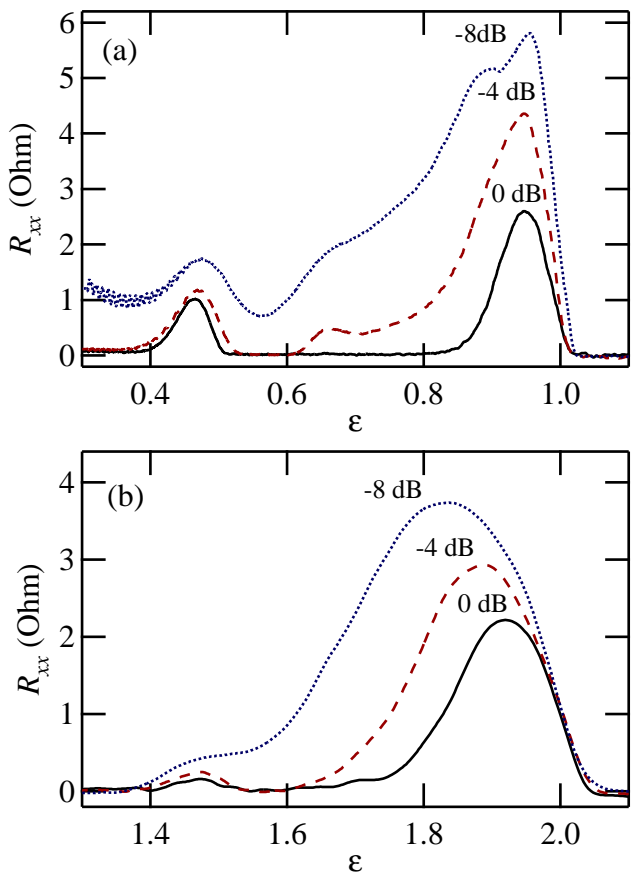}
\caption{[color online]
Magnetoresistance under microwave illumination of $f=27$ GHz as a function of $\varepsilon$, at selected attenuations of -8 dB (dotted [blue] line), -4 dB (dashed [red] line), and 0 dB (solid [black] line), as marked.
Panels (a) and (b) show evolution of photoresistance features at $\varepsilon \approx 1/2,2/3,1$ and $\varepsilon \approx 3/2,2$, respectively.
}
\label{f2}
\end{figure}

In Fig.\,\ref{f1} we show the evolution of photoresistance with increasing MW intensity at $f=\omega/2\pi=27$ GHz.
Here, the intensity is varied by adjusting the attenuator in 2 dB steps starting with -8 dB (top trace) down to 0 dB (bottom trace).
In the top, lowest intensity trace, we easily identify the most prominent two-photon feature corresponding to $\varepsilon_1^{(2)} \approx 1/2$ (see arrow).
As we decrease attenuation to -6 dB another two-photon minimum at $\varepsilon_3^{(2)} \approx 3/2$ (see arrow) appears in magnetoresistance.
The next trace, -4 dB, reveals yet another minimum in the vicinity of $\varepsilon_2^{(3)} \approx 2/3$ (see arrow), which, according to Eq.\,(\ref{mp}), corresponds to a three-photon process.
Most remarkably, we observe for the first time that both two-photon minima now develop into well-resolved ZRS, corresponding to rational $\varepsilon_1^{(2)} \approx 1/2$ and $\varepsilon_3^{(2)} \approx 3/2$.
All observed multiphoton features are accompanied by insets which schematically illustrate the relation between Landau level spacings and the photon energy at corresponding ratios.
Increasing the intensity further, at -2 dB, we observe formation of a three-photon ZRS at $\varepsilon_2^{(3)} \approx 2/3$.
At full microwave power, 0 dB, this three-photon ZRS merges with the strongest two-photon ZRS at $\varepsilon_1^{(2)} \approx 1/2$.
Another interesting feature of the data is the radiation-induced suppression of resistance at $B>0.15$ T (c.f.,  $\varepsilon < 0.4$ in Fig.\,\ref{f2}(a)), where the resistance drops by a factor of 20 with increasing microwave power.

Careful examination of the second-order resistance minima (i.e., at $B\sim 0.028$ T) reveals small negative resistance value ($\sim -0.04$ Ohms), which falls beyond the noise level.
While this value remains unchanged at all microwave intensities studied, it appears to be similar to the features reported by Willett.\cite{willett:026804} 
The origin of this small negative resistance is unclear, but it is unlikely to be caused by the voltage contacts, since, in the same $R_{xx}$ trace clear zeros are observed near $\varepsilon_1^{(1)}$ and $\varepsilon_1^{(2)}$. 

In Fig.\,\ref{f2} we plot the lowest (-8 dB), the middle (-4 dB), and the highest (0 dB) intensity traces as a function of $\varepsilon$.
In panel (a) we concentrate on the regime of $\varepsilon \le 1$, which includes features at $\varepsilon \approx 1/2,2/3,1$, while panel (b) focuses on features at $\varepsilon \approx 3/2,2$.
Plotted without offset, the data reveal that all resistance peaks actually diminish with increasing power.
This strongly non-linear behavior is the characteristic property of the high intensity regime studied here.
With the power increase by roughly a factor of six, the resistance of the first(second) order peak decreases by 55(40)\%, while the two-photon $\varepsilon_{1}^{(2)}\approx 1/2$ peak seems to saturate.
This is in contrast to our earlier lower-intensity, higher-frequency studies\cite{zudov:046807} where the PR maxima increased roughly linearly with MW intensity.
As we already mentioned, the temperature of the coolant was increasing with microwave intensity, from $\approx$\,1.1 K up to $\approx$\,1.8 K.
Observed reduction of the resistance peaks, however, cannot be attributed to the temperature increase alone since at the same time ZRS become stronger.

Another important aspect of the data shown in Fig.\,\ref{f2} is the narrowing of the integral resistance peaks at $\varepsilon_{1}^{(1)},\varepsilon_{2}^{(1)}$ with increasing power.
We notice that in both cases narrowing occurs at the expense of the lower-$\varepsilon$ shoulder, while another edge of the peak is affected only slightly, if at all.
This behavior may be due to multiphoton processes coming into play with increasing power, which induce new rational minima resulting in narrower integral peaks.
However, there is a noticeable difference in the evolution of $\varepsilon_1^{(1)}$ and $\varepsilon_2^{(1)}$ resistance peaks.
While the position of the $\varepsilon_1^{(1)}$ peak does not experience any significant variation, staying at $\varepsilon_1^{(1)} \approx 0.95$, $\varepsilon_2^{(1)}$ peak moves closer to the second harmonic of the cyclotron resonance with increasing power.
In fact, the phase $\phi_2^{(1)}$ of this peak decreases roughly by a factor of two in this range of microwave intensity, from $\approx$\,0.16 at -8 dB down to $\approx$\,0.08 at 0 dB.
With increasing microwave intensity, resistance peaks become more symmetric and it is interesting to compare their widths.
The width of the peaks at the highest microwave power, $\Delta\varepsilon_2^{(1)} \approx 0.158$, $\Delta\varepsilon_1^{(1)} \approx 0.086$, and $\Delta\varepsilon_1^{(2)} \approx 0.056$, increases roughly linear with $\varepsilon$.

\begin{figure}[t]
\includegraphics{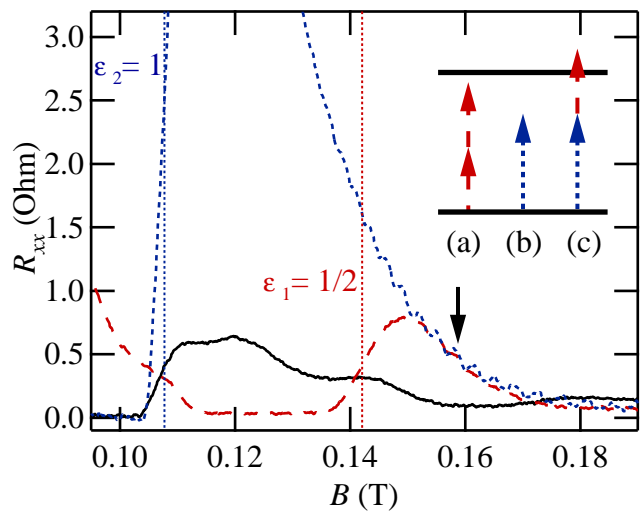}
\caption{[color online]
Magnetoresistance under monochromatic $f_1=31$ GHz (dashed [red] line), $f_2=47$ GHz (dotted [blue] line) and bichromatic, $f_1=31$ Ghz and $f_2=47$ GHz, (solid [black] line).
Vertical dashed [red] and dotted [blue] lines mark $\varepsilon_{\omega_1}^{(2)}=1/2$ and $\varepsilon_{\omega_2}^{(1)}=1$, respectively.
Vertical downward arrow at $\approx$\,0.16 T mark the position where deep bichromatic minimum is formed.
Inset illustrates three possible processes, from left to right, involving (a) two $f_1$ photons, (b) one $f_2$ photon, and (c) one $f_1$ and one $f_2$ photons.
While (a) and (b) processes under-promote an electron and thus correspond to positive photoresistance, (c) results in over-promoting an electron resulting in negative photoresistance.
}
\label{f3}
\end{figure}
Now we turn to the discussion of our bichromatic experiment, where we employ microwave radiation provided by two MW sources of distinct frequencies, $f_1=\omega_1/2\pi=31$ GHz and $f_2=\omega_2/2\pi=47$ GHz.
Detailed results of bichromatic experiments not dealing with multiphoton processes will be published elsewhere.\cite{zudov:up}
The issue that we address here is the bichromatic response of the system in the regime of relatively high magnetic fields, where $\omega_C>\omega_1,\omega_2$.
From the results of monochromatic experiments described above, we expect the formation of rational two-photon peak $\varepsilon_{\omega_1}^{(2)}\approx 1/2$, which is easily identified in Fig.\,\ref{f3} at $\approx$\,0.15 T.
This peak will then overlap with the higher-$B$ tail of the fundamental $\varepsilon_{\omega_2}^{(1)} \approx 1$ peak induced by higher frequency $\omega_2$.
We observe that while the monochromatic resistances almost coincide to the right of the two-photon $\omega_1$ peak, the bichromatic resistance is strongly suppressed.
In fact, observed bichromatic resistance, $R_{xx}^{\omega_1\omega_2}$ is almost one order of magnitude lower than monochromatic resistances $R_{xx}^{\omega_1} \approx R_{xx}^{\omega_2}$, which might indicate formation of a new ZRS.
This new minimum only appears under bichromatic radiation and therefore its origin must be related to a process involving both microwave photons.
We speculate that possible frequency mixing might influence the resonant condition which governs the sign of PR, thus converting PR maximum to a minimum.
For further discussion, we refer to the inset of Fig.\,\ref{f3} where the relation between the cyclotron gap and microwave quanta is graphically presented at $B=0.16$ T.
Both monochromatic $\omega_1$ two-photon (a) and $\omega_2$ one-photon (b) transitions under-promote an electron, and, therefore, correspond to a positive PR.
Adding one of the $\omega_1$ with one $\omega_2$ photon changes the situation and an electron is now over-promoted and produces negative PR, possibly leading to new ZRS.

In summary, we have studied experimentally microwave photoresistance of high-mobility 2DES in the high-intensity regime.
In monochromatic experiments, we have detected new zero-resistance states corresponding to rational values of $\varepsilon$, i.e., $1/2$, $3/2$, and $2/3$, which we associate with two- and three-photon processes.
Formation of these new ZRS with increasing power is complemented by diminishing height, width, and phase of the resistance peaks, as well as dramatic reduction of the resistance at $\varepsilon<0.4$.
Under bichromatic MW radiation we observe strong suppression of resistance appearing at $B\approx 0.16$ T when monochromatic $\varepsilon_{\omega_2}^{(1)}\approx 1$ and $\varepsilon_{\omega_1}^{(2)}\approx 1/2$ peaks overlap.
This suppression possibly originates from a frequency mixing process, $\omega_1+\omega_2$, which alters the resonant condition, and might indicate a formation of a new, bichromatic ZRS.
Taken together, these results demonstrate that multiphoton processes both in monochromatic and bichromatic setups can significantly contribute to the photoresistance of the 2DES and, therefore, should be appropriately addressed in future theoretical and experimental studies.

We would like to thank M. E. Raikh, F. von Oppen, A. Mirlin, and A. Stern for their interest and helpful discussions.
This work is supported by DARPA QuIST and NSF DMR-0408671.


\end{document}